\documentclass[aps, prl, twocolumn, showpacs, footnoteinbib, superscriptaddress, floatfix]{revtex4-1}
\usepackage{graphicx}
\usepackage{amssymb}   
\usepackage{amsfonts}
\usepackage{amsmath}
\usepackage{dsfont}
\usepackage{dcolumn}   
\usepackage{bm}        
\usepackage[mathscr]{eucal}
\usepackage[dvipsnames]{xcolor}
\usepackage[colorlinks, linkcolor=OrangeRed,citecolor=RoyalBlue,urlcolor=NavyBlue]{hyperref}
\usepackage[all]{hypcap} 

\newcommand{\nn}{\nonumber}
\newcommand{\f}{\frac}

\newcommand{\mc}{\mathcal}

\newcommand{\ra}{\rangle}
\newcommand{\la}{\langle}

\newcommand{\eq}[1]{\begin{align}#1\end{align}}

\newcommand{\msr}{\mathscr}

\newcommand{\Pie}{{\Pi}_{\varepsilon}}
\newcommand{\beps}{{\beta}_{\varepsilon}}
\newcommand{\dex}{{\Delta x}}

\begin{document}

\title{Density propagator for many-body localization: finite size effects, transient subdiffusion, and exponential decay}
\author{Soumya Bera}
\affiliation{Department of Physics, Indian Institute of Technology Bombay, Mumbai 400076, India}
\affiliation{Max-Planck-Institut f\"ur Physik komplexer Systeme, N\"othnitzer Stra{\ss}e 38,  01187-Dresden, Germany}
\author{Giuseppe De Tomasi}
\affiliation{Max-Planck-Institut f\"ur Physik komplexer Systeme, N\"othnitzer Stra{\ss}e 38,  01187-Dresden, Germany}
\author{Felix Weiner}
\author{Ferdinand Evers}
\affiliation{ Institute of Theoretical Physics, University of Regensburg, D-93050 Regensburg, Germany}
\begin{abstract}
We investigate charge relaxation in quantum-wires 
of spin-less disordered fermions ($t{-}V$-model).
Our observable is the time-dependent density propagator, 
$\Pi_{\varepsilon}(x,t)$, calculated in windows of 
different energy density, $\varepsilon$,
of the many-body Hamiltonian and at
different disorder strengths, $W$, 
not exceeding the critical value $W_\text{c}$. 
The width $\dex_\varepsilon(t)$ of $\Pie(x,t)$  exhibits a behavior
$d\ln \dex_\varepsilon(t) / d\ln t {=} \beta_\varepsilon(t)$, 
where the exponent function $\beta_\varepsilon(t){\lesssim}1/2$ is seen to depend 
strongly on $L$ at all investigated parameter combinations. 
(i) We confirm the existence of a region in phase space
that exhibits subdiffusive dynamics 
in the sense that $\beps(t){<}1/2$ in large window of times. 
However, subdiffusion might possibly be transient, only, finally giving way 
to a conventional diffusive behavior with $\beps{=}1/2$.  
(ii) We cannot confirm the existence of many-body mobility
edges even in regions of the phase-diagram that have been reported to be 
deep in the delocalized phase.  
(iii) (Transient) subdiffusion $0<\beta_\varepsilon(t)\lesssim 1/2$, 
coexists with an enhanced probability for returning to the origin, 
$\Pie(0,t)$, decaying much slower than $1/\dex_\varepsilon (t)$.  
Correspondingly, the spatial decay of 
$\Pie(x,t)$ is far from Gaussian being exponential or even slower.
On a phenomenological level, our findings are broadly consistent with effects of 
strong disorder and (fractal) Griffiths regions. 
\end{abstract}
\maketitle
{\bf Introduction.}
The discovery of {\em many-body localization} (MBL)
has attracted a considerable attention over recent years and gave 
rise to a new research field~\cite{baa,Mirlin, nandkishore15, 
altman_rev15, MooreRev16}. 
An analytical proof of MBL  has been given with minimal assumptions
 in spin-chains with random local interactions~\cite{Imbrie:2014vo}.  
Such MBL-phases are characterized by the absence of transport 
and  thermalization~\cite{Pal10, Canovi11,luitz15}, which has been  attributed  
to a set of quasi-local integrals of motion~\cite{serbyn13,huse14,chandran14,Ros:2015ib}. 
Anticipating that these integrals of motion adiabatically connect 
to their non-interacting analogues, it is perhaps natural to assume that 
there should be an adiabatic connection between localized 
eigenstates as well~\cite{Bauer:2013jw, Imbrie:2014vo}. 

The MBL-phase is distinguished from another phase
that exhibits a degree of delocalization and which 
therefore is believed to be (thermal) {\em ergodic}~\cite{Pal10,Canovi12,DeLuca:2013ba}. 
The corresponding relaxation dynamics may not, however, 
reflect the simple diffusive behavior familiar from conventional metals.
Instead, a subdiffusive scaling of the \mbox{(spin-)} 
density-correlations has been reported~\cite{BarLev:2015co,
Agarwal:2015cu, barLev16, luitz16, Khait16, 
VipinPRL16}~(though some studies concluded differently~\cite{PrelovsekNoSubPRB16,Brenig15}). 
It was understood to indicate 
Griffiths effects~\cite{vosk15, Agarwal:2015cu, Gopalkrishnan:2015, potter15} near the MBL transition. 
Interestingly, it has been proposed that different behavior 
within these phases may also exist 
that exhibit diffusive relaxation of one 
conserved quantity (charge, energy or spin) and a 
subdiffusive behavior in another~\cite{lerose15, VipinPRL16}.
Clearly, a coexistence of localized and 
delocalized behavior would be incompatible 
with generic expectations based on 
conventional mode-coupling ideas~\cite{Gopalakrishnan:2016}.
\begin{figure}[tb]
\centering
\includegraphics[width=0.85\columnwidth]{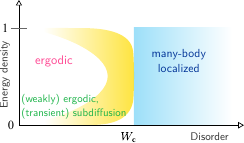}
\caption{A qualitative phase diagram of different dynamical regions in the disorder 
energy-density plane of the $t-V$-model. 
At disorder strength $W$ below the many-body 
localization transition $W_\text{c}$, 
we propose a transient subdiffusive, weakly ergodic dynamical regime
with an anomalously slow decay of the return-probability. 
\label{f1}
}
\end{figure}

The phase transition between the MBL- and the delocalized phase 
is not yet well understood. 
For instance, it has been shown that at very large values of the 
disorder, $W$, {\em all} eigenstates of 
the many-body Hamiltonian $\hat H(W)$ are 
localized~\cite{Znidaric:2008cr,Pal10,luitz15,kjaell14,Aba15}, while
with disorder dropping below a critical value $W{<}W_c$ a transition 
could occur below which $\hat H(W)$ supports a delocalized spectral density window 
~\cite{oganesyan07,kjaell14,luitz15,bera15,vavilov15, LaumanMBME15, Karrasch16} , 
see Fig. \ref{f1}.  
At present, the width of this window is a matter of controversy.
Recent numerical works on the random-field Heisenberg 
chain~\cite{luitz15}, the disordered Ising chain~\cite{kjaell14}, and recent work on Aubry-Andr\'e 
model~\cite{PixleyPRL15} 
were interpreted as giving evidence for the existence of a 
{\em many-body mobility edge} (MBME) 
that separates a band of delocalized states 
from localized band edges. 
%
Later authors have argued, however, that results can be significantly 
contaminated with finite size effects unless carefully extrapolated. For instance, the  
phase-boundary as found in Ref.~\cite{luitz15} 
should be shifted to large disorder values as argued in Ref.~\cite{Devakul15}. 
In fact, the very existence of MBME was called in question  
by De Roeck {\it et al.}, who suggested that the presence 
of a delocalized spectral window should imply the possibility 
for the formation of hot bubbles of electronic liquid 
that destabilize localizing processes in all spectral density windows~\cite{deRoeck16}.  

In this work, we investigate the charge propagation focussing on 
the delocalized region near the MBL transition. A common description 
of relaxation dynamics employs the density propagator, $\Pi(x,t)$, 
that takes a simple Gaussian 
shape for diffusive systems:  
$\Pi(x,t){=}e^{-\frac{1}{2}(x/\Delta x(t))^2}/\sqrt{2\pi}\Delta x(t), 
\Delta x(t){=}\sqrt{Dt}$, where $D$ is the diffusion constant. 
Aiming at mobility edges, we actually study a variant of 
it, $\Pie(x,t)$, that resolves the contribution to $\Pi(x,t)$ 
stemming from many-body states with energy densities $\varepsilon$.
We thus get access to the 
length scales relevant for the crossover physics, which allow us to carefully monitor finite 
size and finite time effects. In this way we go  beyond previous studies.  

We outline our results: (i) Within our observation window, $\Pie(x,t)$
exhibits a very pronounced non-Gaussian spatial shape that decays in 
a (simple) exponential fashion or even slower. It is tempting to associate 
this finding with the stretched exponential behavior 
of correlations that has recently been proposed to exist 
due to fractal Griffiths regions in 
the localized phase near the phase boundary~\cite{zhang16}.
(ii) Due to this peculiar shape of $\Pie(x, t)$, the time dependence of its width 
$\Delta x_\varepsilon(t)$, is very 
sensitive to the system size, $L$. In order to highlight the effects of finite size in the time 
evolution, we investigate the exponent scaling function
\begin{equation}
\label{e1}
\beta_\varepsilon(t) \equiv \frac{d \log \Delta x_\varepsilon(t)}{d\log t}, 
\end{equation}
which at long times quantifies the rate of growth of $\Delta x_\varepsilon(t) \propto 
t^{\beps(t=\infty)}$ and for diffusive systems $\beps=1/2$.
In the ergodic phase at intermediate times $\beps(t)$ grows in a subdiffusive manner 
with values $\beta_\varepsilon(t) {<} 1/2$ consistent with the earlier 
reports~\cite{BarLev:2015co,Agarwal:2015cu,luitz16,Khait16, VipinPRL16}. 
However with increasing time, $\beps(t)$ 
becomes progressively $L$-dependent.
At these longer times a similar tendency of growing $\beps(t)$ 
(with $L$) is observed in all spectral windows 
-- at low, intermediate and high energy density. 
This strong growths prevents us from confirming 
the existence of genuine subdiffusion 
that would exibit a time-independent exponent $\beps{<}1/2$.  
We detect a slow growth of $\beps(t)$ 
even in those regions of the phase diagram that have been identified previously as 
localized. Thus, 
the (delocalized) phase is larger than reported previously, 
which is associated with a very slow collective dynamics.
\footnote{Following a recent proposal, such a behavior is not 
entirely unexpected, perhaps signalizing the breakdown of localization due to 
``hot bubbles''~\cite{deRoeck16}.}

(iii) For the probability $\Pie(0,t)$ to return to the origin
one might have suspected 
$\Pie(0,t) {\propto} 1/{\dex_\varepsilon(t)}$, suggesting
$\Pie(0,t)\propto t^{-\beps(t=\infty)}$. 
Instead, our data indicates that the subdiffusive 
transients coexist with  an elevated 
return probability consistent with (possibly transient) 
 weakly ergodic sub-phases with fractal phenomenology, 
$\Pie(0,t) \propto \dex_\varepsilon(t)^{-\alpha_\varepsilon}$ 
and  $0\leq \alpha_\varepsilon<1$. 


{\bf Model and Method.} 
Like several works before~\cite{oganesyan07, Berkelbach:2010ib,DeLuca:2013ba, luitz15, bera15, 
BarLev:2015co, BeraConc16,gdtMBL16}, 
we consider the $t{-}V$-model  
\eq{
\mc{\hat{H}}  = & -\frac{t_\text{h}}{2}\sum_{x=-L/2}^{L/2-2} \hat{c}^\dagger_{x} \hat{c}_{x+1} + h.c.  + 
\sum_{x=-L/2}^{L/2-1}  \mu_x  \left ( \hat{n}_x - \frac{1}{2}\right) \nn \\
 &  + V\sum_{x=-L/2}^{L/2-2}   \left ( \hat{n}_x - \frac{1}{2}\right)   \left ( \hat{n}_{x+1} - 
\frac{1}{2}\right), 
\label{eq:H}
}
where the summations are along an $L$-site wire, $x{=}1,\ldots,L$, 
with hopping~($t_\text{h}=1$)
and interaction~($V$) between nearest neighbors, only; 
the uncorrelated on-site energies $\mu_x$ 
are being drawn from a box distribution $\lbrack-W,W\rbrack$. 
We work at a half filling and with open boundary conditions. For $V=1.0$, the MBL transition is 
believed to be at $W_\text{c}\approx 3.5$~\cite{luitz15}. 
The specific correlator $\Pie(x, t)$ that we are interested in has not
yet been investigated; it is defined via its Fourier space 
representation~\footnote{Our definition of the discrete 
Fourier transform of $x_n$: 
$y_q = \sum_{n=0}^{L-1} x_n e^{-\imath q n}$, $q{=}\frac{2 \pi 
a j}{L}$ and lattice spacing $a{=}1$.}: 
\begin{equation}
\Pi_\varepsilon(q,t) = 
\overline{\Phi_\varepsilon (q,t)/\Phi_\varepsilon (q,t{=}0^{+})}, 
\end{equation}
where the disorder average is denoted by the overline. 
$\Phi_\varepsilon (q,t)$ is the Fourier transform of the energy-projected 
density relaxation functions
\begin{eqnarray}
\label{e3}
 \Phi_\varepsilon(x,t)
 &=& 
 \left[ \langle \hat n_x(t)\hat n_0\rangle_\varepsilon 
 - \langle \hat n_x\rangle_\varepsilon \langle \hat n_0\rangle_\varepsilon 
 \right] \Theta(t) \text{.} 
 \label{e9}
\end{eqnarray}
The projection into a narrow spectral range 
near $\varepsilon$ is facilitated by taking the 
expectation value of an operator
$
\langle \hat {\mathcal O} \rangle_\varepsilon
=\text{Tr} \hat {\mathcal O} \hat \rho(\varepsilon)
$
with 
\eq{
\hat{\rho}({\varepsilon}) = \mc{N}^{-1} 
\int_{\varepsilon-\Delta\varepsilon/2}^{\varepsilon+\Delta\varepsilon/2} d\varepsilon' 
\sum_{\gamma}^{\mc{N}} |\gamma \ra
\delta(\varepsilon_\gamma-\varepsilon')\la \gamma | 
\text{,}
}
where $|\gamma \ra$ denotes the  eigenstates of the Hamiltonian~\eqref{eq:H}  
with energy-density $\varepsilon_\gamma{=}(E_\gamma{-}E_\text{min})/(E_\text{max}{-}E_\text{min})$, where 
$E_\gamma$ are the many-body energies and $E_\text{max, min}$ 
denote the extremal values of the 
energy spectrum. 
$\mc{N}$ represents the number of states 
in the energy density window $\Delta \varepsilon$, and it is exponentially large in $L$. 
By definition,
$\Pi_\varepsilon(q{=}0,t){=}1$
and for a conventional diffusive system we have a 
Gaussian shape, 
$\Pi_\varepsilon(q,t){=}\exp(-(\Delta x_\varepsilon(t)q)^2)\Theta(t)$, 
with $\Delta x_\varepsilon(t)=\sqrt{D_\varepsilon t}$. 
For the time evolution, Eq.~\eqref{e3}, we employ
a standard Chebyshev-polynomial propagation ~\cite{Wei06}; 
traces over operators are performed stochastically as averages 
over random state vectors. 
The approach owes its efficiency to the fact that disorder 
averages converge very rapidly with the number of random 
states. 
Details of the calculations and performance tests we 
relegate to the supplementary material. 

\begin{figure*}[tb]
\centering
\includegraphics[width=1\textwidth,keepaspectratio=true]{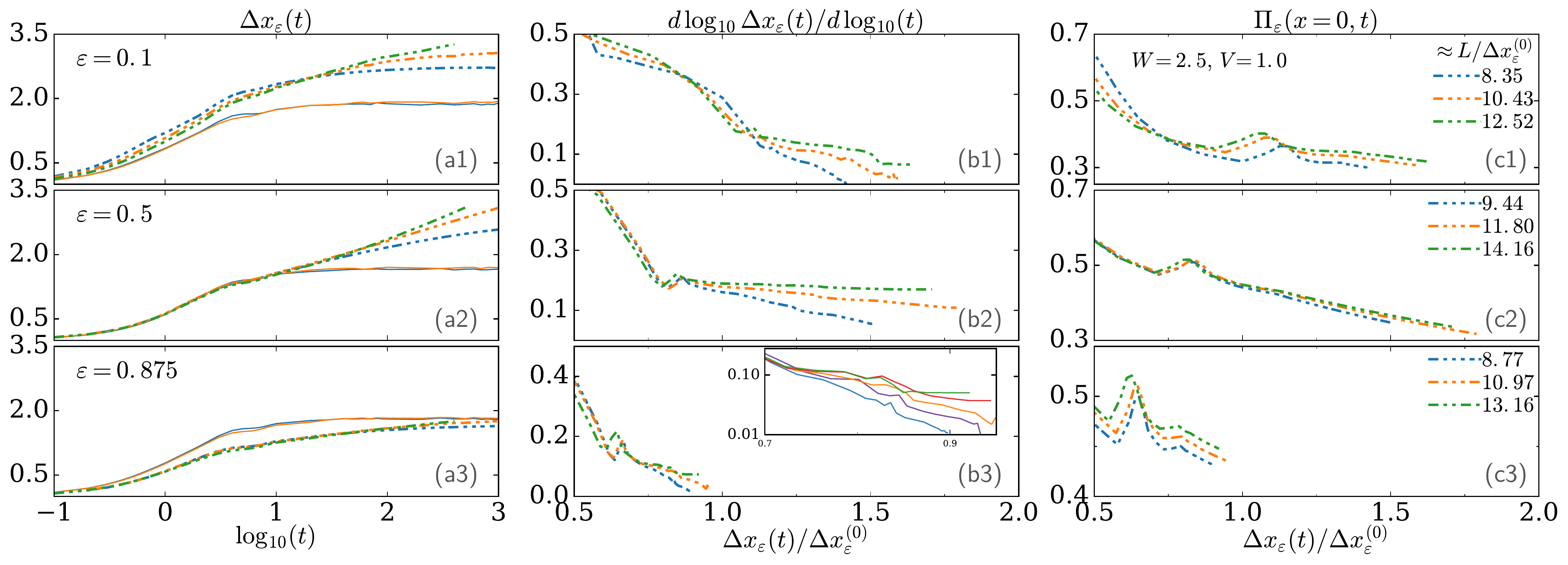}
\caption[Evolution of the width $\xi$]
{(a1)-(a3)~The time evolution of $\dex_\varepsilon(t)$ at $W{=}2.5$ 
and $V{=1}$ near the 
lower band-edge (upper row,~$\varepsilon=0.1$) in the center region (center row,~$0.5$) 
and near the upper band-edge (lower row,~$0.875$) for system sizes $L=16,20,24$
(dashed traces blue, red, green). 
Also shown are 
non-interacting reference traces 
for $L{=}16,20$ ($V{=}0$, solid lines).
%
%
(b1)-(b3)~Re-plotting (a1)-(a3) as 
$d\ln \Delta x_\varepsilon(t)/d\ln t$ over $\Delta x_\varepsilon(t)/\Delta x_\varepsilon^{(0)}$
to highlight finite-size effects. 
Inset shows the blow up of the (b3) data for better visibility of trends
including system sizes $L=16,18,20,22,24$~(bottom to top). 
(c1)-(c3)~Probability to return to the origin.
The legends in this column also give the 
three system sizes in units
of the bare localization length. 
(In all calculation we fix the width of the energy window
$\Delta \varepsilon=0.1$~\cite{supp}.) 
\label{f2}}
\end{figure*}
{\bf Results.} 
We begin the analysis of the propagator $\Pie(x, t)$ 
with its second moment in real space, 
$$
\dex_\varepsilon(t)^2 {=} \langle x^2\rangle_\varepsilon - \langle  
x\rangle_\varepsilon^2 \text{,} \quad \langle x^n\rangle_\varepsilon 
= \sum\limits_{x=-L/2}^{L/2-1}  x^n \ \Pie(x,t).
$$
%
Fig.~\ref{f2}-(a1-a3) show the $\dex_\varepsilon(t)$ at $W =2.5$ for both interacting~($V 
= 1$, dashed line) and non-interacting ($V = 0$, solid line) case
for several values of energy densities ($\varepsilon=0.1,0.5,0.875$).
For these parameters MBMEs have been reported 
near $\varepsilon\approx0.2$ and near $0.8$
with a delocalized regime in between~\cite{luitz15}. 

Figure~\ref{f2}-(a1-a3) carries several messages. 
(i) Finite size effects are very strong: the system size, $L$, 
exceeds the non-interacting standard deviation, $\dex_\varepsilon^{(0)}$~(saturation value in time), 
by a factor of 10-15~($\approx L/\dex_\varepsilon^{(0)}$), but 
nevertheless the growth of $\dex_\varepsilon(t)$ changes with $L$ by as much as 30\%. 
(ii) The interaction mediated delocalization process is very slow. 
Even after a time that typically corresponds to 0.1\% of the inverse hopping 
$t^{-1}_\text{h}$
the width of the wavepacket has grown by less than a factor of 
two as compared to $\dex_\varepsilon^{(0)}$. 
(iii) Depending on the spectral window, the transient dynamics is 
quite different. In particular, the spreading of $\Pie(x,t)$ 
is enhanced by the interactions at low energy densities while it is 
hindered at high densities as compared to the non-interacting reference case.


\begin{figure*}[tb]
\centering
 \includegraphics[width=1\textwidth,keepaspectratio=true]{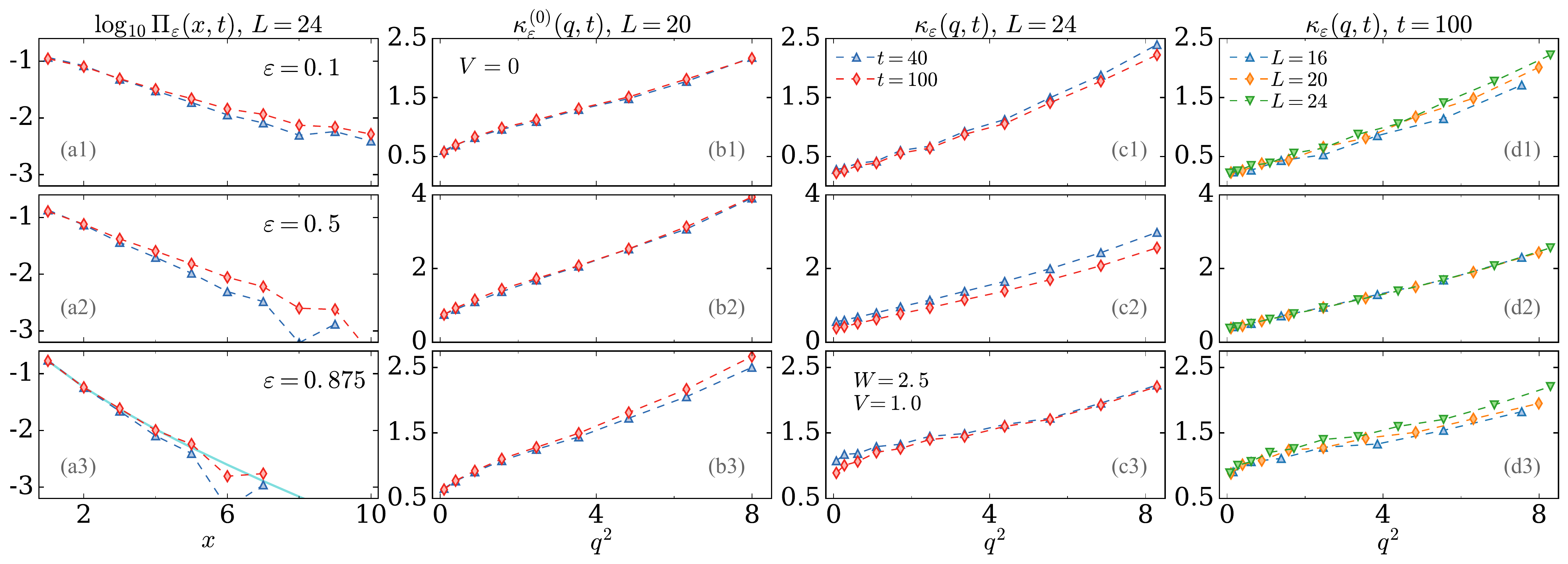}
\caption[Density propagator]
{
(a1)-(a3)~The density propagator $\Pie(x,t)$ in the delocalized regime 
($\varepsilon{=}0.1,0.5,0.875$, $W{=}2.5$, $L{=}24$)
at two times $t{=}40, 100$. 
The log-normal plot illustrates non-Gaussian shape. Solid line in (a3) shows a stretched exponential fit with an exponent $\approx 0.7$. 
(b1)-(d3) The corresponding memory kernel 
$\kappa_\varepsilon(q,t){=} q^2/(\Pie^{-1}(q,t)-1)$, 
see also \eqref{e7}, for the case without (b1)-(b3)
and with interactions (c1)-(d3). 
The structure at larger wavenumbers
illustrates the (non-exponential) short-distance behavior.
The absence of effects in time (and system size, not shown)
highlights the localized character of the non-interacting 
kernel $\kappa_\varepsilon^{(0)}$. 
In contrast, the evolution of the interacting kernel
is the hallmark of delocalization. 
(d1)-(d3)~Shows the $L$-dependence of $\kappa_\varepsilon(t)$. 
\label{f3}}
\end{figure*}
{\bf Flowing exponent -- $\beps(t)$.}
To quantify the time dependence of $\dex_\varepsilon(t)$, we study the $\beps(t)$ as 
defined in Eq.~\eqref{e1}.
Fig.~\ref{f2}-(b1-b3) shows the $\beps-$function as a function of 
$\dex_\varepsilon(t)/\dex_\varepsilon^{(0)}$.
It very clearly highlights the fact
that beyond a certain transient time, $\tau_\varepsilon$~(set by the kink position),
a slow dynamics sets in which reveals itself by a 
high degree of sensitivity to the system size, $L$.
Moreover, as is seen in Fig.~\ref{f2}-(b1-b3) all traces of $\beps(t)$ experience a kink 
with a position evolving with the energy density $\varepsilon$
that does not collapse after rescaling of the abscissa with $\dex_\varepsilon^{(0)}$.

While the range of $L$-values available to us is not 
sufficient to study the asymptotic limit (in $L$ and $t$), 
our data nevertheless gives a non-vanishing lower bound for 
$\beps(t)$ and hence indicates delocalization, 
at least near the band-center. 
With this caveat, we notice 
that the qualitative behavior seen in all energy ranges
is the same: With $L$ increasing, there is a pronounced trend for 
$\beps(t)$ to grow (at fixed long time), see Fig.~\ref{f2}-(b1-b3) 
and inset. 
%
Strictly speaking, we thus find no evidence for an 
upper bound to $\beps$ below the diffusion limit $1/2$, 
i.e. for genuine subdiffusion.
Moreover, the growth (with $L$) being similar in all energy windows, 
we also find no evidence for the existence of a 
many-body mobility edge at $W{=}2.5$. 
The picture is similar for other choices of $W~(\lesssim 3.0)$~\cite{supp}. 
At larger disorder and close to the transition, $W{\approx}W_\text{c}$, 
the situation is numerically less conclusive due to residual statistical noise. 
To account for this in Fig.~\ref{f1}, 
this region of the phase-diagram has been left uncolored (white).

{\bf Return probability -- $\Pie(x{=}0,t)$.}
In one dimensional diffusive systems the return probability 
associated with a spreading 
wavepacket relates to the variance $\Pie(0,t)\sim 1/\dex_\varepsilon(t)$, 
merely stating that the wavepacket is internally homogeneous. 
The data displayed in Fig.~\ref{f2}-(c1 - c3) does not adhere 
to this fundamental idea: $\Pie(0,t)$ is close to stationary and 
therefore does not follow the $1/\dex_\varepsilon$ law, 
most clearly seen in the low and high energy density regimes.
This observation finds a natural explanation adopting the idea  
of strong disorder induced fractality. 
%
%
Indeed it is well known that in the presence of 
(multi-)fractality the return-probability can be enhanced, 
 $\Pi_\varepsilon(0,t)\propto \dex_\varepsilon^{-\alpha_\varepsilon}$, 
with $0\leq \alpha_\varepsilon < 1$~\cite{ketzmerick97}.  
A very slowly decaying return probability
can therefore also indicate a fractal-type behavior, 
i.e. $\alpha_\varepsilon$ being significantly smaller than unity. 
Unfortunately, it is very challenging to extract $\alpha_\varepsilon$
reliably from our data, because our observation window for $\dex_\varepsilon(t)/\dex_\varepsilon^{(0)}$ does not exceed 
a factor 2-3. 
 
%

{\bf Density propagator -- $\Pie(x,t)$.}
To understand the transient sub-diffusive behavior further, here we look at the time dependence of 
the full distribution function, $\Pie(x,t)$, both in real and $q$-space.  
Fig.~\ref{f3}-(a1-a3) displays a density-propagator $\Pie(x,t)$ that 
is far from Gaussian. 
To highlight its shape~(curvature at small $q$,large $x$) we rewrite $\Pi_\varepsilon(q,t)$ 
employing an (inverse) memory kernel, $\kappa_\varepsilon(q, t)$, 
\begin{equation}
\label{e7}
\Pi_\varepsilon(q,t) = \left(1+q^2/\kappa_\varepsilon(q,t)\right)^{-1}, 
\end{equation}
where $ -\partial_q^2 \Pi_\varepsilon(q,t) \rvert_{q=0} = 2/\kappa_\varepsilon(0,t) 
\sim\dex_\varepsilon(t)^2$. 
A numerical example can be read off from Fig.~\ref{f3}-(b-c).
It displays $\kappa_\varepsilon$ at three different 
energy densities at intermediate disorder strength $W{=}2.5$.
%
%
Notice that the non-interacting kernel, 
$\kappa_\varepsilon^{(0)}(q,t)$,  
is rapidly growing with wavenumber, $q$~(see Fig.~\ref{f3}(b1-b3)). 
This behavior reflects the presence of a short-distance cutoff, 
$a$, such as the lattice constant, terminating the
long-distance, exponential tail. It exists in a similar way also 
in the interacting kernels $\kappa_\varepsilon(q,t)$, 
see Fig.~\ref{f3}-(c-d)~\footnote{Notice that 
$\kappa_\varepsilon$ in Fig.~\ref{f3}-(b1,b3) 
exhibits small oscillations in $q$ that result from the finite system size.}~\footnote{We would 
like to draw attention to a small additional feature 
that emerges for the high-energy kernel at very 
small wavenumbers; as seen in Fig.~\ref{f3}-(c3) 
with increasing time a cusp develops.
It could be seen as a precursor indicating a stretched exponential 
shape in real space and the corresponding fit is shown in Fig.~\ref{f3}-(a3).  
Its emergence at high-energies first is understandable 
because of the relatively weak tendency to delocalization signalized 
by the observation $\dex_\varepsilon(t)<\dex_\varepsilon^{(0)}$.}.

{\bf Conclusions.}
In this work, we have considered the full space-time structure of the 
spectrally resolved density correlator, $\Pie(x,t)$,  allowing us 
to monitor finite size effects. 
%
%
(i) The processes that are characteristic of delocalized behavior are very slow. 
Even at observation times of order $10^3$ 
(in units of inverse hopping $t_\text{h}^{-1}$), 
 $\Pie(x,t)$ has spread over little more than the 
non-interacting length, $\dex_\varepsilon^{(0)}$. 
(ii) Although the system size exceeds $\dex_\varepsilon^{(0)}$ by a large factor, 
finite size effects are substantial reflecting a spreading of  
$\Pie(x,t)$ that is far from Gaussian, possibly (stretched) exponential in the tails.

Because of strong finite-size effects, the exponents $\beps(t)$ 
that describe the spreading dynamics of the variance of 
the density propagator, $d\ln \dex_\varepsilon(t)/d\ln t = \beps(t)$, 
are hard to quantify reliably. 
We are able to provide a lower bound for $\beps(t)$ suggesting 
the absence of many-body mobility gaps in the $t-V$-model at values of $W$ not too close 
to the transition region -- apparently consistent with recent analytical arguments~\cite{deRoeck16}.
Since we cannot provide an upper bound for $\beps(t) < 1/2$, 
we cannot confirm the existence of genuine  
subdiffusive behavior 
in the asymptotic
limit; a logically possible alternative is a transient behavior with an
effectively growing exponent $\beps(t)$ that gradually converges to
the diffusion limit $1/2$. 
Together with {\em transient} subdiffusive behavior, we observe a drastically enhanced return 
probability, which could be interpreted as $\Pi_\varepsilon(0,t)\propto 
\dex_\varepsilon^{-\alpha_\varepsilon}$ in accord with the assumptions of fractality
induced by strong-disorder physics.

Based on these findings we propose the following scenario: 
There is a timescale $\tau_\varepsilon$ beyond which 
a slow dynamics kicks in together with diffusive behavior. 
Approaching the MBL transition from the delocalized side, this time scale diverges;
simultaneously, $\beps(t)$ at times $t{\lesssim} \tau_\varepsilon$ is rapidly
decreasing, which might suggest a small value 
of $\beps$ at the MBL transition. 
In this scenario, the critical fixed-point would carry 
excited states that exhibit phenomenological features 
reminiscent of (strong) multifractality~\cite{evers08}.

%
We conclude with two remarks relating our work to the most 
recent literature. 
(a) Consistent with our findings, also Serbyn et. al.
observe very strong finite size effects in their study of the 
Thouless energy~\cite{serbyn16}. 
Like us, they interpret their results as indicating 
that the system sizes are too short for observing 
the asymptotic thermalized behavior. 
Unlike us, they go a step further proposing that 
the numerical data at small system sizes (below $L{=}20$) 
already reveals hydrodynamic properties 
of the critical fixed point, such as multifractality.
This conclusion for us is difficult to draw, because 
one would expect system-size independent exponents in the 
critical window, which we don't observe.  
%
%
(b) Recent studies of Anderson localization of
random regular graphs (RRG) reveal a slow flow 
with system size out of a \mbox{(quasi-)}multifractal into an ergodic
regime~\cite{GarciaMata16,Tikhonov16}.
When interpreting $\dex_\varepsilon(t)$ as an effective system size, then 
the transient subdiffusive behavior observed by us finds 
a natural interpretation within the RRG-perspective. 


{\bf Acknowledgments.}  Discussions with I. Gornyi, A. D. Mirlin and D. Polyakov are gratefully 
acknowledged. SB and GDT also thank M. Heyl for discussions. 
The project was supported by DFG under projects EV30/7-1
and EV30/11-1. SB acknowledges support from the 
ERC starting grant QUANTMATT NO. 679722.
\bibliography{references}

\setcounter{equation}{0}
\setcounter{figure}{0}
\setcounter{table}{0}
\makeatletter
\renewcommand{\theequation}{S\arabic{equation}}
\renewcommand{\thefigure}{S\arabic{figure}}
\renewcommand{\bibnumfmt}[1]{[S#1]}
\renewcommand{\citenumfont}[1]{S#1}
\newpage
\makeatother
\clearpage 
\onecolumngrid
\appendix

\appendix{{\large \bf Additional material for `Density propagator for many-body localization: finite size effects, transient 
subdiffusion, and exponential decay'}}

\section{Validation of the numerical method}
The energy-projected density relaxation function is the main object studied in this work. 
It is defined as
\eq{
 \Phi_\varepsilon(x,t) = 
 \left[ \langle \hat n_x(t)\hat n_0\rangle_\varepsilon 
 - \langle \hat n_x\rangle_\varepsilon \langle \hat n_0\rangle_\varepsilon 
 \right] \Theta(t) \text{,} 
 \label{seq:phi_e}
}
where $\langle \hat {\mathcal O} \rangle_\varepsilon=\text{Tr} \hat {\mathcal O} \hat \rho(\varepsilon)$, and 
$\hat \rho(\varepsilon)$ projects into a narrow
spectral range near energy density $\varepsilon$ with width $\Delta\varepsilon$. 
To calculate the two-point space-time correlator 
\eqref{seq:phi_e} for large systems~($L{=}24$) 
and long times~(${\approx}10^3$), we use two approximations: (i) The
energy
projected trace denoted via the angular brackets 
$\langle \ldots \rangle_\varepsilon$ is evaluated stochastically, 
while (ii) the time evolution is performed employing a standard kernel-polynomial 
method based on Chebyshev polynomials. 
In this section, we detail and validate (i) and (ii), including  
data illustrating the convergence properties. 
\subsection{Chebyshev-representation of the density matrix $\hat{\rho}({\varepsilon})$}
\begin{figure}[tbh]
 \includegraphics[width=0.5\columnwidth]{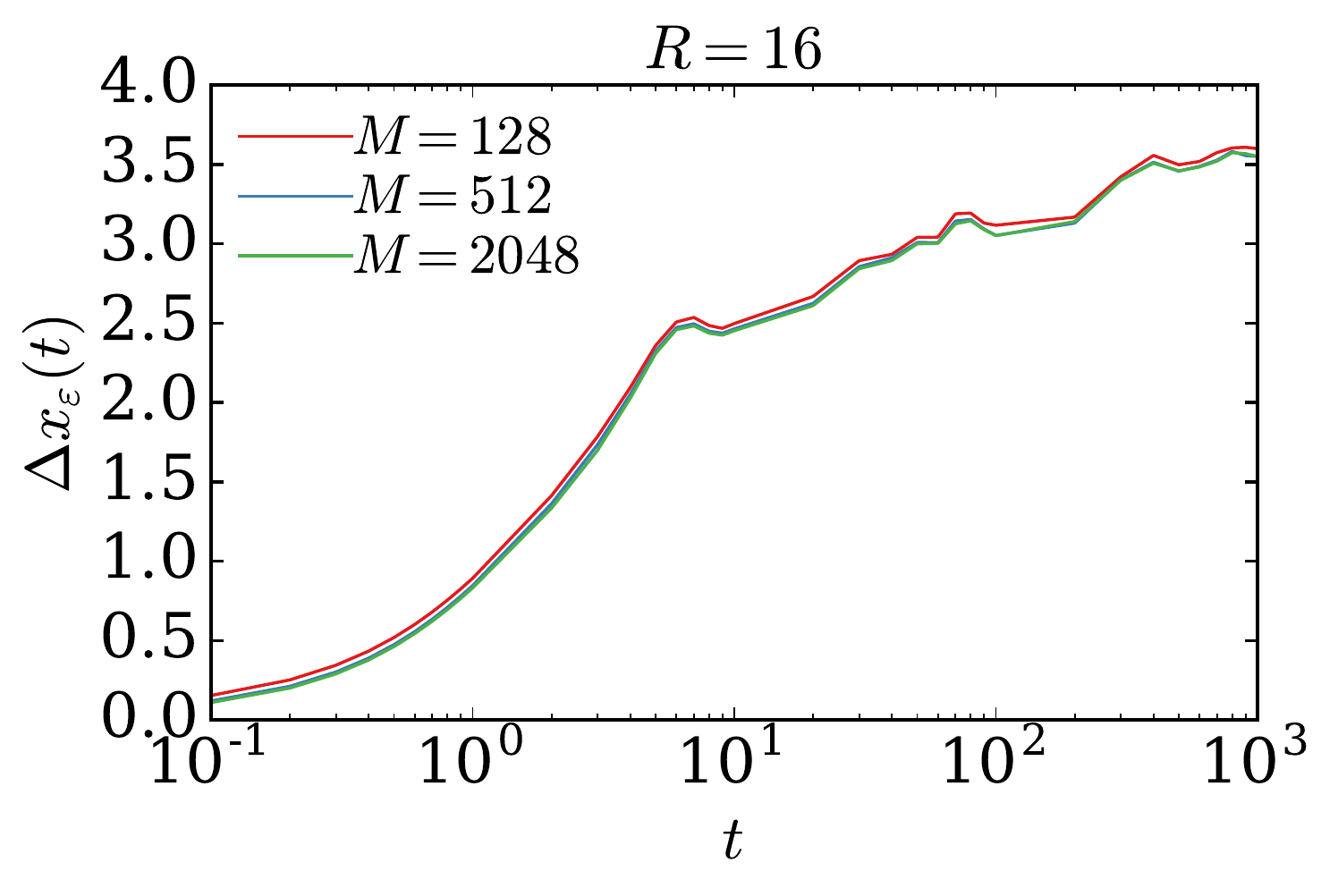}
 \caption{
  Convergence of the variance of the 
density propagator
(defined explicitly in the main paper) with respect to 
the number of moments, $M$, 
used in the expansion~\eqref{seq:rho2}. $R$ defines the number of random vectors taken for the trace evolution~(see text for 
definition). Only 16 
disorder samples is taken for averaging. 
 }
\label{sf:rho}
\end{figure}

For numerical evaluation we represent the density matrix 
$\hat{\rho}({\varepsilon})$ as a simple function of the 
Hamiltonian $\mc {\tilde H}$ (and is rescaled between energy density $\{0,1\}$) in the following 
way,  
\eq{
 \hat \rho (\varepsilon)= \frac{\msr{R}_{ [ \varepsilon-\Delta\varepsilon/2,  \varepsilon+\Delta\varepsilon/2 ] }( 
\mc {\tilde H}) }{ \text{Tr} \, \msr{R}_{ [ \varepsilon-\Delta\varepsilon/2,  
\varepsilon+\Delta\varepsilon/2 ] }( \mc 
{\tilde H})} , 
\label{seq:rho}
}
where $\msr{R}_{[a,b]}(x)$ is the box function of unit height in the interval $[a,b]$. 
We approximate $\hat{\rho}({\varepsilon})$ as a truncated Chebyshev series, 
\eq{ 
\hat{\rho}({\varepsilon}) \approx \frac{\sum_{i=0}^{M}\mu_i T_i(\mc {\tilde H} ) }{\text{Tr} 
\sum_{i=0}^{M}\mu_i T_i( \mc 
{\tilde H}) } 
\label{seq:rho2}
}
where $\{T_i (x)\}$ denote the Chebyshev polynomials. 
$M$ denotes the order of the expansion taken sufficiently 
large~($M\geq 3000$) to assure convergence \eqref{seq:rho2} 
(see also Fig.~\ref{sf:rho} right panel). 
The expansion coefficients $\{\mu_i\}$ 
are given as follows: $\mu_0 = \frac{1}{\pi} ( \arccos{a} -  \arccos{b})$,  $\mu_1 = 
\frac{1}{\pi} ( \sqrt{1-a^2} - \sqrt{1-b^2})$, $\mu_{n\ge 2} = \frac{1}{n\pi} ( \sin{(\arccos{nb})}  
- 
\sin{(\arccos{na})})$. 
%

In Fig.~\ref{sf:rho} we display the
convergence of the time evolution 
of our main observable, 
the variance $\Delta x_\varepsilon(t)$, 
with respect to the number of moments in the sum
\eqref{seq:rho2}. For definition of the $\Delta x_\varepsilon(t)$, see main text. 

\subsection{Stochastic trace evaluation and convergence}
\begin{figure}[htb]
 \includegraphics[width=1\textwidth]{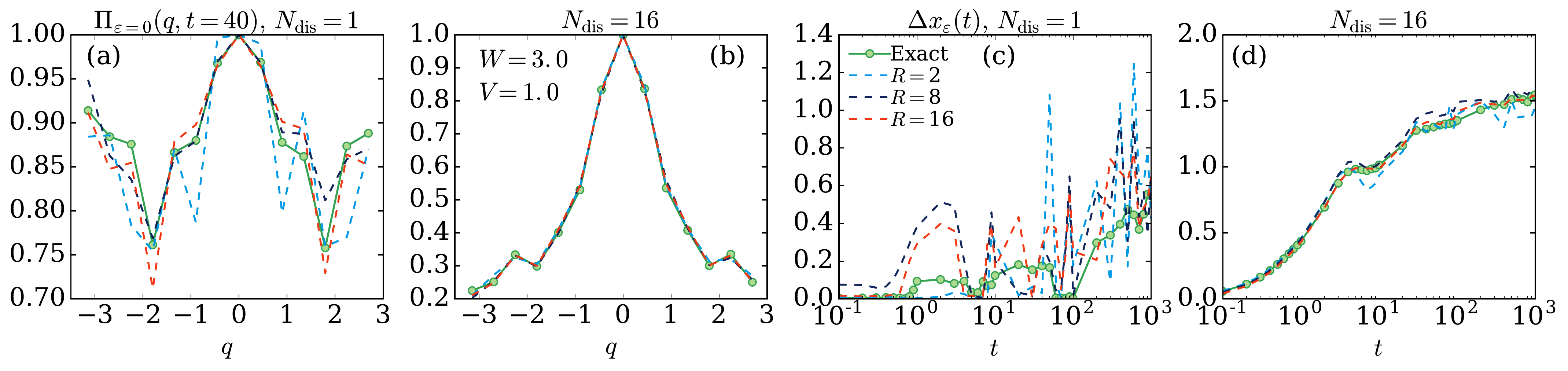}
 \caption{Trace evaluation: comparison between exact and stochastic methods 
 for $\Pie(q,t)$ and $\Delta x_\varepsilon(t)$). 
 (Parameters: $L=14$, middle of the band $\varepsilon=0.5$ and $W=3.0, V=1.0$). 
 (a) Density propagator $\Pie(q, t=40)$ in $q$-space for a single disorder 
 realization. 
 The (green) dots represent the exact data calculated using the full trace employing exact 
diagonalization; the dashed lines are evaluated with different 
number of random vectors $R=\{2, 8, 16\}$ 
(blue, black, red)
employing the stochastic trace 
formula~\eqref{seq:trace}. 
(b) Density propagator $\Pie(q,t)$ averaged over 16 disorder realizations. 
As can be seen, the average of $\Pie(q,t)$ over the disorder realization converges rapidly 
in the number $R$ of stochastic state vectors as opposed to $\Pie(q,t)$ taken for a single 
disorder realization. 
(c),(d) A similar trend 
is also visible with real space data, here shown for the second moment 
of $\Pie(x,t)$: $\la \Delta x^2(t) \ra$.}
\label{sf:tarce_exact}
\end{figure}

The expectation values  $\langle \hat {\mc O} \rangle_\varepsilon$ 
have been calculated using \textit{stochastic trace evaluation}. 
The idea is to represent a trace as an average over an 
ensemble of random state vectors $\{ |r \rangle \}_{r=0}^R$:
\eq{ 
\langle \hat {\mathcal O} \rangle_\varepsilon \sim  \frac{1}{R} 
\sum_{r=0}^{R-1} \langle r |
\hat{\rho}({\varepsilon}) \hat {\mathcal O} |r 
\rangle, \quad \text{with}\; R\gg1.
\label{seq:trace}
}
\begin{figure*}[htb]
 \includegraphics[width=0.8\textwidth]{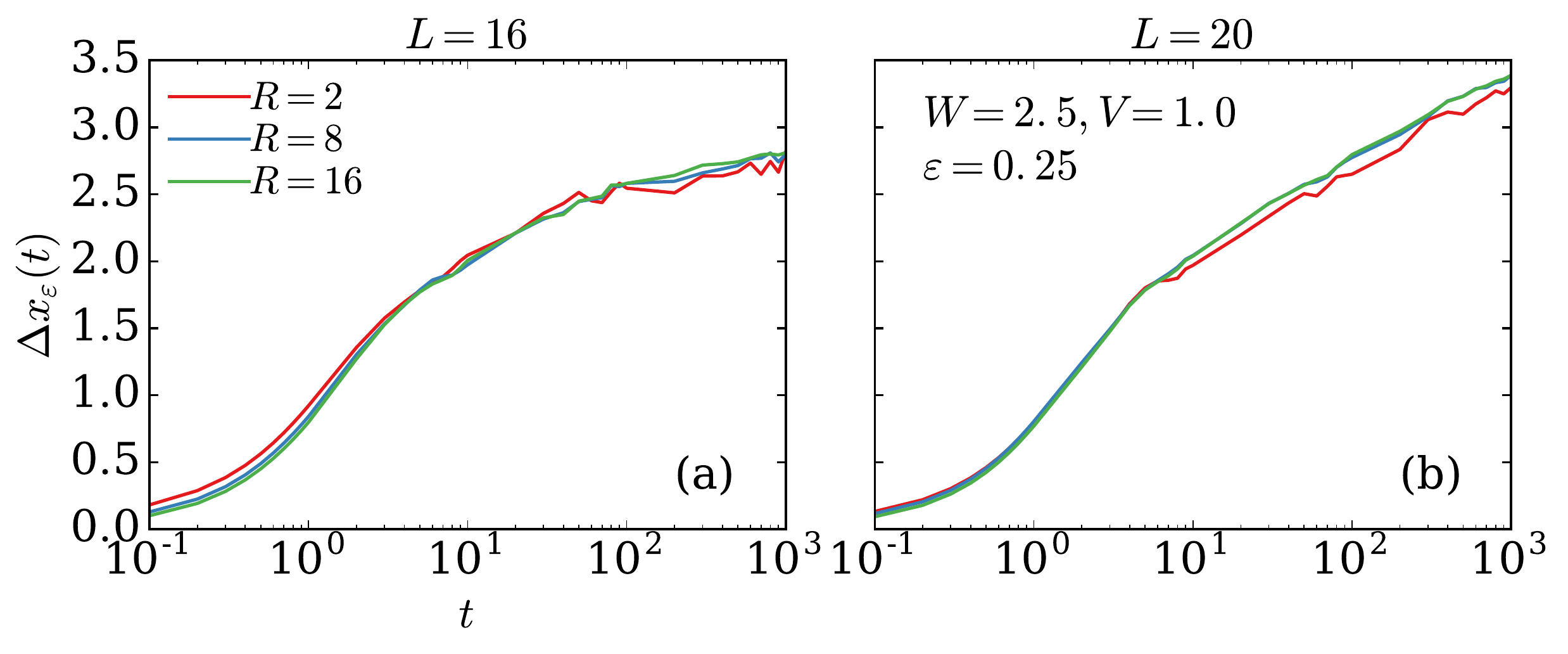}
 \caption{Shows the variance $\Delta x_\varepsilon(t)$ for three trace vectors $R=\{2, 8, 16\}$ after a small disorder averaging 
$N_{\text dis}=32$ 
for two different system sizes $L=16, 20$.}
\label{sf:tarce_vecs}
\end{figure*}
Truncating the sum at an upper cutoff, $R$, 
for global variables the relative error decays as  
$1/\sqrt{D R}$,  
 $D$ denoting the dimension of the Hilbert space.
Hence, the stochastic trace evaluation is most efficient 
in very high dimensions (for variables that sample the full system size). 
In our case, $D$ is exponentially large in the system size, $L$, 
and is given by  ${L \choose N}$, $N$ being the particle number. 
For smaller system size, $L \lesssim 20$, 
we typically use $R{=}16$ random state vectors, while for
larger system sizes we only keep $R{=}2$.
The convergence properties are illustrated in 
Fig.~\ref{sf:tarce_exact}-(a). 
The plot displays a comparison between the stochastic trace estimate 
and an exact trace evaluation. As is seen there, the
convergence properties of the distribution $\Pie(q,t)$ with $R$ 
are actually quite poor; at $R{=}16$ deviations are still 
of the order of a few percent.

however, note that the convergence with $R$ is drastically improved for 
the traces averaged over the disorder ensemble, 
i.e. for $\overline{\langle \hat {\mc O} \rangle_\varepsilon}$  
Fig.~\ref{sf:tarce_exact}-(b) shows that even for a relatively 
small ensemble of $N_\text{dis}{=}16$ samples a good convergence 
is reached already with $R{=}2$.
The same behavior is seen at all times. To illustrate this we 
display similar data also for the variance, $\Delta x_\varepsilon(t)$. 
Again, the disorder averaged variance converges very rapidly with 
the number $R$ of random states kept for the trace evaluations. 

Figure~\ref{sf:tarce_vecs} further illustrates 
the dependence of the variance on the averaging over trace vectors, 
now for two larger system sizes. 
As is obvious from
both  plots, the variance is 
well approximated at all times with only a small number of trace vectors. 
With increasing system size and improving disorder average the trace 
approximation becomes progressively efficient. 
This is because the error scales as $\propto 1/\sqrt{D}$, 
where $D$ is the dimension of the Hilbert space, 
which increases exponentially fast with the system size $L$.

\begin{figure*}[htb]
 \includegraphics[width=1\textwidth]{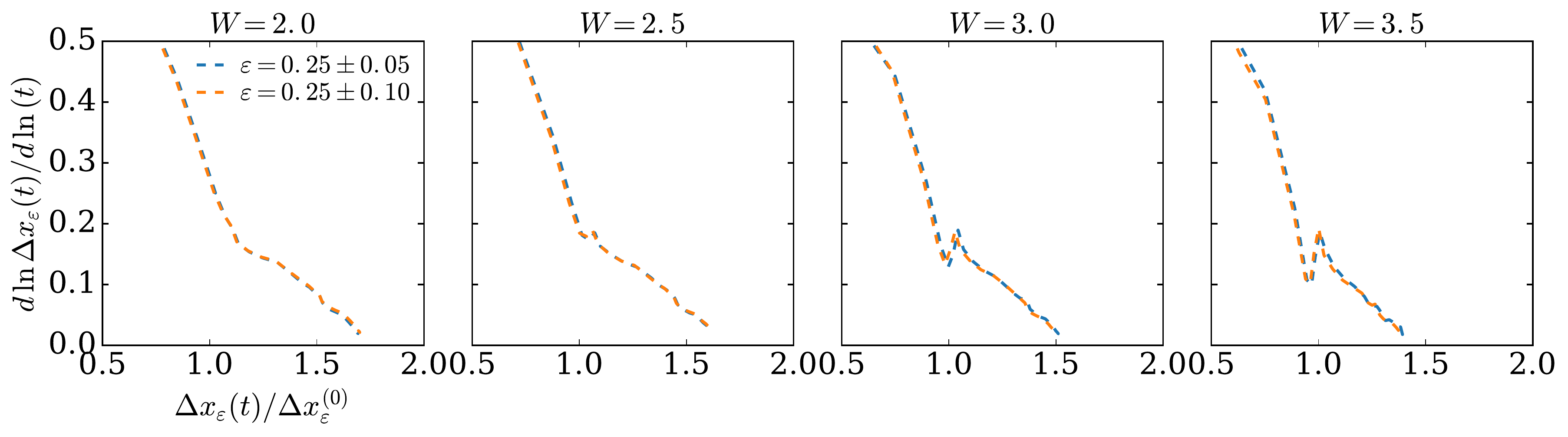}
 \caption{Evolution of the exponent 
 $\beta_\varepsilon(t) = d\ln \Delta x_\varepsilon(t) / d\ln t $ for
different values of the width of the energy density $\Delta \varepsilon$ 
for $L=16$ and disorder strengths
$W=\{2.0,2.5,3.0,3.5\}$ and $\varepsilon=0.25$.}
\label{sf:diff_epsilon}
\end{figure*}
\subsection{Time evolution: Chebyshev expansion}
The time evolution of the operators has been preformed relying  
 once more the standard Kernel polynomial techniques~\cite{KPM}
employing Chebyshev expansions of the 
exponential of the $\mc{\hat {H}}$:  
  \eq{
    U(t) \approx e^{-i b t} \sum_{k=0}^{N} \mu_k T_k (\tilde{\mc{H}}); \qquad \mu_k = (-i)^k J_k(at), 
    \label{seq:Ut}
    }
where $\tilde{\mc{H}} =  \f{\mc{\hat H}{-}b}{a}$ denotes the rescaled 
Hamiltonian; $a=(E_\text{max} {-} E_\text{min} )/2$, 
$b=(E_\text{max} {+} E_\text{min} )/2$ are the scaling factors
and $J_k(x)$ denotes the Bessel function of order $k$. We
typically take $N \gtrsim 2 a t$ 
to ensure convergence~\cite{Weise2008} of the truncated Chebyshev series~($T_k(x)$).
Eq.~\eqref{seq:Ut} only requires sparse matrix multiplications. 
The iterative scheme scales as $\mc{O}(M)$ as compare to
exact diagonalization which is $\mc{O}(M^3)$, 
$M$ denoting the dimension of $\mc{\hat H}$. 
Therefore, system sizes up to $L=24$ can be treated 
for times of the order $\approx 10^3$ 
(in units of inverse hopping $t_\text{h}=1.0$). 


\subsection{Dependence on $\Delta\varepsilon$ }
We have ascertained that our choice of the width
$\Delta\varepsilon$ of the energy-density shell 
 was sufficiently narrow so that our
results for $\Pie(x,t)$ and its variance 
are (essentially) independent of it. 
In Fig.~\ref{sf:diff_epsilon} we show the evolution of the 
exponent $\beta_\varepsilon(t)$ for two different values of
the width $\Delta\varepsilon=0.1, 0.2$ of the box 
function \eqref{seq:rho} at energy density $\varepsilon=0.25$. 
The data is averaged over $\gtrsim10^4$ disorder configurations. 
As is easily deferred from the 
figure, the curves are almost indistinguishable form each other. 
In all the data presented in the main part of the work  
we choose $\Delta \varepsilon=0.1$. 

\section{Results: Dependence of $\beta_\varepsilon(t)$ on disorder and energy density}
\begin{figure*}[b]
 \includegraphics[width=1\textwidth]{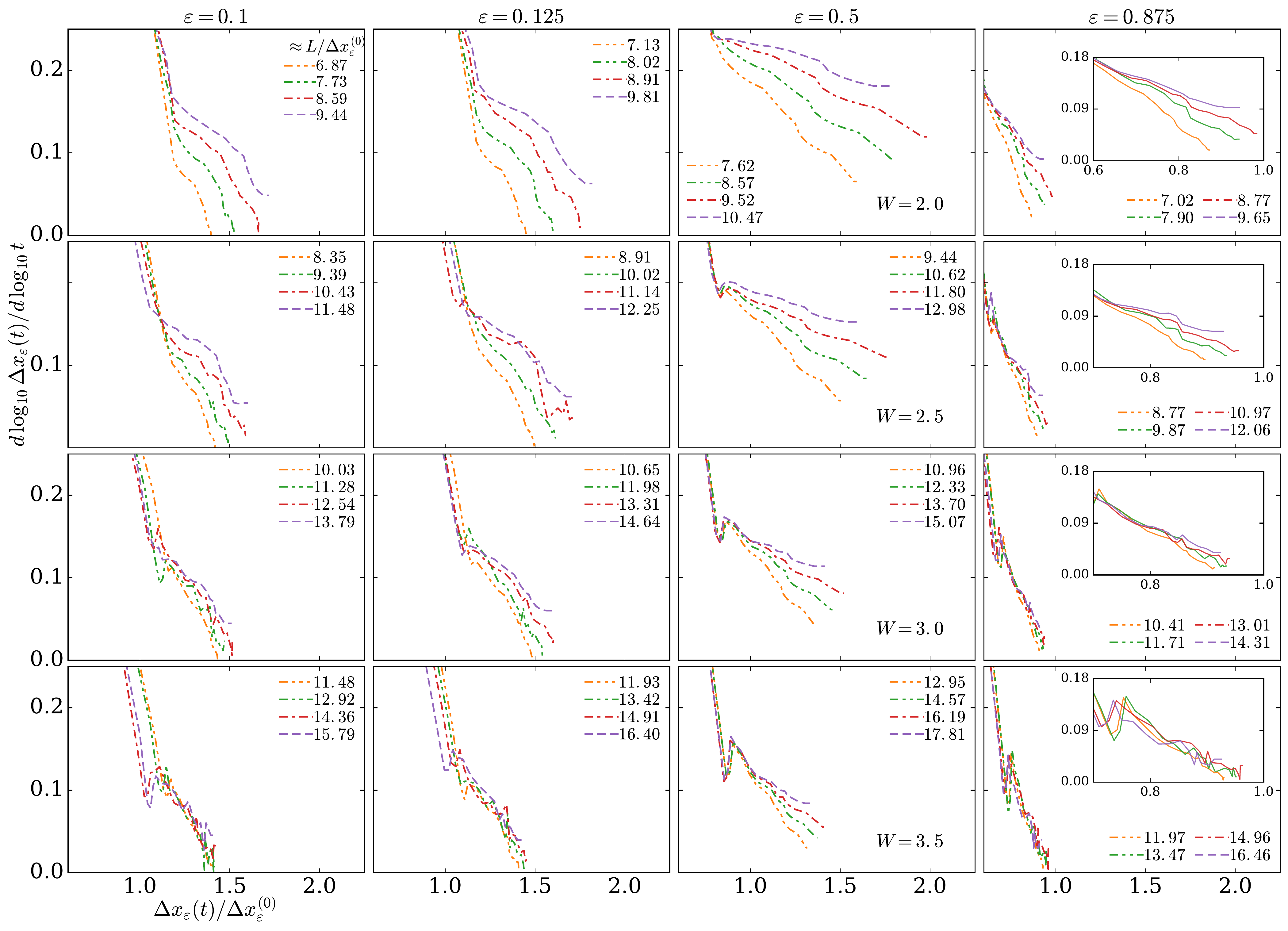}
 \caption{Time dependence of the exponent 
 $\beta_\varepsilon(t) = d\ln \Delta x_\varepsilon(t) / d\ln t $ for
different disorder $W=\{2.0,2.5,3.0,3.5\}$ 
and system sizes ($L{=}16,18,20,22$) at four different energy densities 
$\varepsilon=\{0.1, 0.125, 0.5, 0.875\}$
with $\Delta \varepsilon=0.1$ and $V=1.0$. Inset: 
Shows the same data as fourth column but zoomed for better 
visibility of the trend in the data 
with increasing system sizes.}
\label{sf:diff_beta}
\end{figure*}
Fig.~\ref{sf:diff_beta} shows the evolution of the $\beta_\varepsilon(t)$ over $\Delta 
x_\varepsilon(t)/\Delta
x_\varepsilon^{(0)}$ for 
$L=\{16,18,20,22\}$, at four energy densities and four  
disorder values close to the many-body localization transition 
($W=\{2.0, 2.5, 3.0, 3.5\}$), which is believed to be around $W_\text{c}\approx 3.5$.  For these 
data we usually perform around $10^6$ disorder 
realizations for small system sizes~($L \lesssim 20$), while for larger system sizes the data 
is averaged over around $10^4$ disorder samples.

\paragraph{$\varepsilon{=}0.5$, }  Fig.~\ref{sf:diff_beta} (3rd column):
We start our discussion from the middle of the spectrum. 
In this regime the data clearly indicates that the dynamics is 
(transient) subdiffusive with an (effective) exponent, 
$\beta_\varepsilon(t) < 1/2$, 
which depends strongly on system size $L$. 
The $L-$dependence is reflected via the upward movement of the
$\beta_\varepsilon(t)$. We interpret this systematic trend
as an indication to delocalization. 
\paragraph{$\varepsilon=0.1, 0.125$:} 
Fig.~\ref{sf:diff_beta} (1st, 2nd column) 
shows the evolution of the exponent $\beta_\varepsilon(t)$ for different systems sizes
in the low energy-density regime. 
Previous studies assigned this region to the many-body localized phase 
(at $W\gtrsim 2.0$, $V=1.0$). 
However, for disorder strength below $W \lesssim 3.5$, the upward trend seen with these 
curves is similar to one in the band center thus suggesting the presence of a 
(slow) delocalization mechanism which is inconsistent with the assignment to the MBL-phase
and the existence of a mobility gap in this parameter range. 
The proliferation of statistical noise precludes a further analysis 
about whether or not at even larger disorder, $W\approx 3.5$, an MBME could exist. 
The noise enhancement near the spectral edges simply reflects the low spectral weight
and thus is not unexpected. 
\paragraph{$\varepsilon=0.875$:} 
Statistical noise and finite size effect are largest in the 
high energy-density regime~(Fig.~\ref{sf:diff_beta}, 4th column). 
At disorder values below $W\lesssim 3.0$ a systematic 
delocalizing trend at largest times is seen, which also here we would like 
to interpret as an indication of a very slow delocalization mechanism. 
Concerning statements about MBME at larger disorder values, we consider our data
to be inconclusive due to strong statistical fluctuations. 

\begin{figure*}[b]
 \includegraphics[width=1\textwidth]{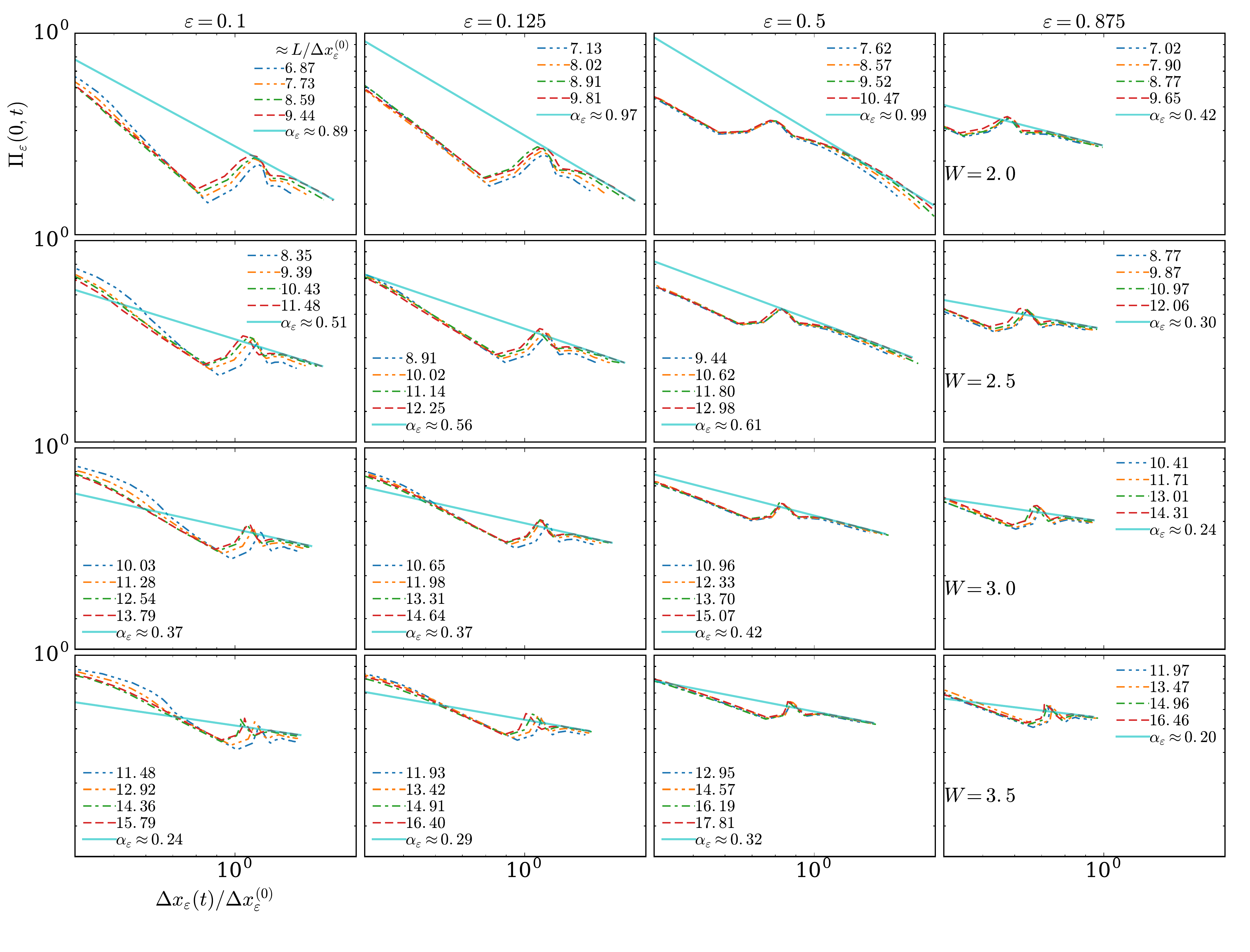}
 \caption{Time dependence of the return probability  
 $\Pi_\varepsilon(0, t)$ in double log scale for
different disorder $W=\{2.0,2.5,3.0,3.5\}$ 
and system sizes ($L{=}16,18,20,22$) at four different energy densities 
$\varepsilon=\{0.1, 0.125, 0.5, 0.875\}$
with $\Delta \varepsilon=0.1$ and $V=1.0$. The solid line serves as a guide of a power-law fit and 
also an estimate of the corresponding exponent $\alpha_\varepsilon$ is provided. }
\label{sf:diff_rp}
\end{figure*}

\section{Return probability $\Pi_\varepsilon(x=0,t)$}
Fig.~\ref{sf:diff_rp} shows the evolution of the $\Pi_\varepsilon(0,t)$ over $\Delta 
x_\varepsilon(t)/\Delta
x_\varepsilon^{(0)}$ for 
$L=\{16,18,20,22\}$, at four energy densities and four  
disorder values close to the many-body localization transition 
($W=\{2.0, 2.5, 3.0, 3.5\}$). The slow decay of the return probability is clearly visible  
for disorder values not too far from the transition. A power law fit of the data is also provided 
to highlight the slowness of the decay. However, due to the small time window (only a factor of 2 
in $\Delta x_\varepsilon(t) /\Delta x_\varepsilon^{(0)}$) the fit is not completely reliable and 
should be taken only as a 
guide to eye.

\section{Testing a stretched exponential decay}
\begin{figure}[tbh]
 \includegraphics[width=0.5\textwidth]{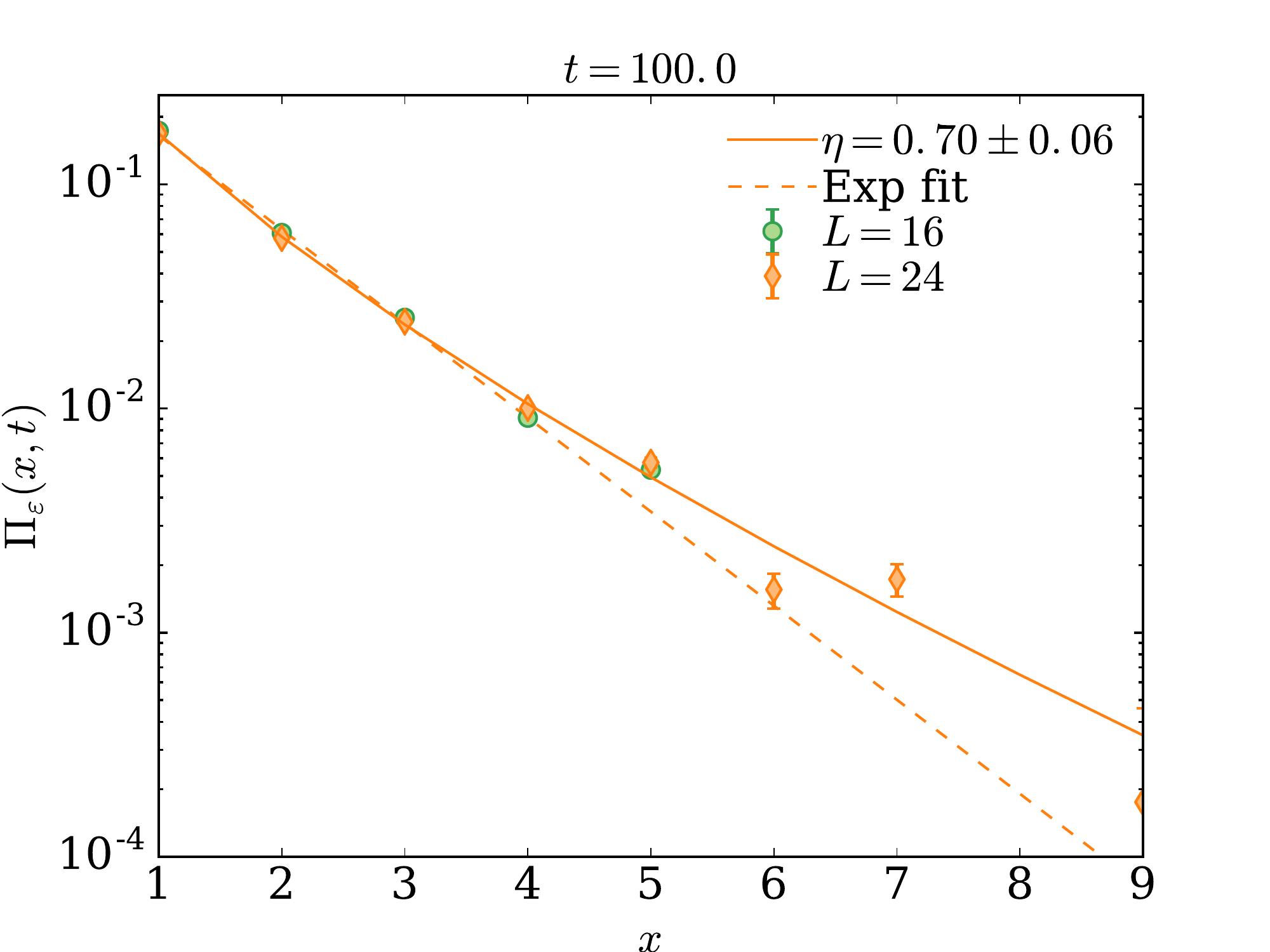}
 \caption{Distribution function $\Pie(x,t)$ in real space exhibiting a 
 decay slower than exponential in the tail region. 
 Solid line represents a stretched exponential fit, 
 $\exp(-(x/\xi)^\eta)$,  with fitting parameters 
 $\eta\approx 0.7, \xi{=}0.464\pm0.12$. For comparison, 
 the dotted line indicates a simple exponential.
 (Parameters: $\varepsilon=0.875$, $L=24$, $W=2.5,V=1.0$ at an intermediate time $t=100$.) 
 We have also shown the corresponding data for $L{=}16$ (green symbols) 
 to ascertain that finite-size effects are negligible. 
  }
\label{sf:s_alpha}
\end{figure}

Fig.~\ref{sf:s_alpha} shows the distribution function $\Pie(x,t)$ in real space taken 
in the subdiffusive phase at high energy density in the vicinity of the 
MBL-transition. In the tail region a weak upturn is seen that indicates deviations 
from a simple exponential behavior. We describe the data on a phenomenological level 
employing a stretched exponential, three parameter fit 
$\Pie(x, t) \approx \exp(-|x/\xi|^\eta)$. 
Indeed, the fitting suggests that the exponent $\eta$ is significantly smaller 
than one, $\eta \approx 0.7$. 
\end{document}